# The Rise of Big Bang Models, from Myth to Theory and Observations

## Jean-Pierre Luminet

*Laboratoire Univers et Théories, CNRS-UMR 8102,*
*Observatoire de Paris, F-92195 Meudon cedex, France*
*jean-pierre.luminet@obspm.fr*

**Abstract.** We provide an epistemological analysis of the developments of relativistic cosmology from 1917 to 2006, based on the seminal articles by Einstein, de Sitter, Friedmann, Lemaitre, Hubble, Gamow and other main historical figures of the field. It appears that most of the ingredients of the present-day standard cosmological model, such as the acceleation of the expansion due to a repulsive dark energy, the interpretation of the cosmological constant as vacuum energy or the possible non-trivial topology of space, had been anticipated by Lemaitre, although his papers remain desperately unquoted.

## *1. From Myth to Science*

What are the origins of the universe, of the stars, of the earth, of life, of man? These questions have given rise to many different myths and legends, and today they are more than ever the subject of intensive research by astrophysicists, biologists and anthropologists. What were once fanciful stories are now scientific models ; but, whatever form they take, ideas about the origins of the universe both reflect and enrich the imagination of the people who generate them. Every society has developed its own stories to explain the creation of the world ; all of them are ancient myths rooted in religion.

Whereas in monotheistic religions God is believed to have existed before the Creation, in most other kinds of religion the gods themselves are thought to originate from a creative element such as Desire, the Tree of the Universe, the Mundane Egg, Water, Chaos or the Void.

A belief in some such primordial element, of which there are traces in every culture, underlies man's thinking about the history of the cosmos like a primitive universal symbol buried in the collective subconscious. This may explain the vague links which can always be discerned between this or that creation myth and modern scientific descriptions of the origin of the universe –for example, big bang theory. There is therefore nothing mysterious or surprising about these correspondences other than that certain ways of thinking about the world should be so ingrained in the human mind.

In fact scientific and mythical explanations of the origins are neither complementary not contradictory; they have different purposes and are subject to different constraints. Mythical stories are a way of preserving collective memories which can be verified neither by the story-teller nor by the listener. Their function is not to explain what happened at the beginning of the world but to establish the basis of social or religious order, to impart a set of moral values. Myths can also be interpreted in many different ways. Science, on the other hand, aims to discover what really happened in historical terms by means of theories supported by observation. Often considered to be anti-myth, science has in fact created new stories about the origin of the universe: big bang model, the theory of evolution, the ancestry of mankind. It is therefore hardly surprising that the new creation stories developed by scientists tend to be regarded by the general public as modern myths.

Relativistic cosmology, which is now the accepted frarnework within which both the structure and the evolution of the universe are described, is full of examples of this kind of mythologising, despite the fact that it is less than 100 years old. Indeed the ideological basis on which Einstein built his 1917 model of a static, eternal universe was partly philosophical; in order to complete the structure, he had to invent a factor called the « cosmological constant » and incorporate it into his general theory of relativity. The discovery that the universe was expanding meant that Einstein's model had to be abandoned and in 1931, Georges Lemaître proposed a scientific explanation of the birth of space and time, according to which the universe resulted from the fragmentation of a « primeval atom » – a theory which may recall the ancient notion of everything hatching from a Mundane Egg. Lemaître's model was subsequently revised and adjusted to become the basis of big bang theory.

The concept of « continuous creation », which enjoyed some popularity in the 1950's, is an even more striking example of scientific myth making. At the time, big bang theory had not yet been fully substantiated by observation and many astrophysicists were reluctant to accept its metaphysical implications. Among them were Hermann Bondi and Thomas Gold, who in 1948 put forward the « steady state » theory, whose fundamental principle, known as the « perfect cosmological principle », related back to Aristotle. The Greek philosopher had maintained that the world was eternal and indestructible and therefore without beginning or end, in contradiction of Plato, who in *Timaeus* had expressed the idea that the world had begun and would end at a specific time. To compensate for the gradual dilution of matter that would result from the constant expansion of the universe, the steady state theorists had to invoke the idea of continuous creation of matter, at the rate of about one atom of hydrogen per cubic meter of space every five billion years. In 1948 the English astronomer Fred Hoyle demonstrated that the steady state model was feasible on condition that a new field (which he called simply C for « creation ») was added to the equation; this *ad hoc* invention was envisaged as a reservoir of negative energy which had existed throughout the life of the universe – *i.e.* for ever. The idea of continuous creation had appeared many times before (in the legends of the Aztecs, who believed that constant human sacrifices were necessary to regenerate the cosmos, as well as in the writings of Heraclitus and the Stoics, for example) and the scientific theory clearly followed this tradition. The theorists, however, had to « force » their model to fit their philosophical views by introducing unrealistic processes. The discovery of the cosmic background radiation in 1965 finally disproved their hypothesis and provided evidence for Lemaître's big bang theory.

## 2. The beginnings of Relativistic Cosmology : static solutions.

The History of Relativistic Cosmology can be divided into 6 periods :

– the *initial* one (1917-1927), during which the first relativistic universe models were derived in the absence of any cosmological observation.

– a period of *development* (1927-1945), during which the cosmological redshifts were discovered and interpreted in the framework of dynamical Friedmann-Lemaître solutions, whose geometrical and mathematical aspects were investigated in more details.

– a period of *consolidation* (1945-1965), during which primordial nucleosynthesis of light elements and fossil radiation were predicted.

– a period of *acceptation* (1965-1980), during which the big bang theory triumphed over the « rival » steady state theory.

– a period of *enlargement* (1980-1998), when high energy physics and quantum effects were introduced for describing the early universe.

– the present period of *high precision experimental cosmology*, where the fundamental cosmological parameters are now measured with a precision of a few %, and new problematics arise (nature of the dark energy, topology of the universe, new cosmologies in quantum gravity theories, etc.)

Let us follow chonologically the rather hectic evolution of the ideas in the field.

In 1915, Einstein (and also Hilbert) provided correct field equations for a relativistic theory of gravitation. This new theory, coined as « General Relativity », proposed a new frame of reference for understanding the universe. Einstein argued that gravity was not a force but an effect of the curvature of space-time caused by the distribution of mass and energy. The theory allows scientists to overhaul cosmological theory : the universe, until then a relatively vague concept, acquires a new consistency and becomes a physical entity defined by its space-time strcuture and its composition of matter, light, radiation, in fact all kinds of energy. Space-time is endowed with a rich structure, expressed geometrically in terms of curvature and topology, and physically in terms of its matter and energy contents.

Cosmological solutions of Einstein's equations are obtained by assuming homogeneity and isotropy in the matter-energy distribution. This implies that space curvature is on the average constant (*i.e.* it does not vary from point to point, although it may change with time).

The first exact solution was obtained in 1917 by Einstein himself (see Luminet 2004 for detailed references), who quite naturally wished to use his brand new theory of general relativity to describe the structure of the Universe as a whole. He assumed that space has a positive curvature, *i.e.* the geometry of the Riemannian hypersphere, and searched for a static solution, *i.e.* in which the average matter density is constant over time, as well as the radius of the hyperspherical space. Einstein expected that general relativity would support this view. However this was not the case. The model universe that he initially calculated did not have a constant radius of curvature : the inexorable force of gravity, acting on each celestial body, had a tendency to make it collapse. For Einstein, the only remedy was to add an *ad hoc* but mathematically coherent term to his original equations. This addition corresponds to some sort of « antigravity », which acts like a repulsive force that only makes itself felt at the cosmic scale. Thanks to this mathematical trick, Einstein's model remained as permanent and invariable as the apparent Universe. The new term, called the « cosmological constant », has to keep exactly the same value in space and time. Formally, it can take any value whatsoever a priori, but Einstein fitted it to a specific value $\lambda_E$ in order to constrain the radius $R_E$ of the hypersphere and the matter density $\rho$ to remain constant over time. He thus derived the relation $\lambda_E = 1/R_E^2$ .

For Einstein, the fact that space had to be static was a natural assumption since at this time, no astronomical observation indicated that stars have large velocities. In fact, his main motivation was to get a *finite* space (although without a boundary) and try to fit his solution with Mach's ideas about the origin of inertia : « I must not fail to mention that a theoretical argument can be adduced in favour of the hypothesis of a finite universe. The general theory of relativity teaches that the inertia of a given body is greater as there are more ponderable masses in proximity to it; thus it seems very natural to reduce the total effect of inertia of a body to action and reaction between it and the other bodies in the universe, as indeed, ever since Newton's time, gravity has been completely reduced to action and reaction between bodies. From the equations of the general theory of relativity it can be deduced that this total reduction of inertia to reciprocal action between masses - as required by E. Mach, for example - is possible only if the universe is spatially finite. »[1]

---

[1] A. Einstein, *Geometry and Experience*, Address to the Prussian Academy of Sciences in Berlin on January 27[th], 1921.

In the same year, 1917, the Dutch astrophysicist Willem de Sitter derived another model for a static relativistic universe, which was very different from that of Einstein. He assumed that space is positively curved (in fact the projective hypersphere, also called elliptic space, where the antipodal points of the ordinary hypersphere are identified), and empty (in other words, the matter density is zero). As a counterpart, in the absence of matter and therefore of gravity, only a cosmological constant can determine the curvature of space, through the relation $\lambda = 3/R^2$. A strange consequence was that, although the hyperspherical space was assumed to be static (i.e. $R$ = const), the spatial separation between any two test particles (stars, galaxies) had to increase with time. This meant that the cosmological constant has a particular influence on the structure of space : it generates « motion without matter ». For Einstein, de Sitter's solution reduced to a simple mathematical curiosity, since the real Universe indeed has a mass.

## 3. *The rise of dynamical solutions*

In an article which appeared in 1922, entitled *On the Curvature of Space* (see Luminet 2004 for reference and translation), the Russian physicist Alexander Friedmann took the step which Einstein had balked at : he abandoned the theory of a static universe, proposing a « dynamic » alternative in which space varied with time. As he stated in the introduction, « the goal of this notice is the proof of the possibility of a universe whose spatial curvature is constant with respect to the three spatial coordinates and depend on time, e.g. on the fourth coordinate. ».

Thus he assumed a positively curved space (hypersphere), a time variable matter density $\rho(t)$, and a vanishing cosmological contant. He obtained his famous « closed universe model », with a dynamics of expansion – contraction.

For the first time in the history of cosmology, the problem of the beginning and the end of the universe was couched in purely scientific terms. Friedmann also derived solutions with non zero cosmological constant, but pointed out that the term was superfluous. Contrarily to a current opinion, Friedmann's work was not purely mathematical ; but he was honest enough to recognize that the available astronomical observations could not support his model : « Our information is completely insufficient to carry out numerical calculations and to distinguish which world our universe is. [...] If we set $\lambda = 0$ and M = 5 × $10^{21}$ solar masses, the world period becomes of the order 10 billion years. » It was a remarkable prediction, since the most recent estimate for the age of the universe is about 14 billion years.

Friedmann was not only a brilliant physicist, he was also a fervent orthodox catholic : for him, general relativity suggested creation of the world by God (although he did not formulated this statement in a published work).

Einstein reacted quickly to Friedmann's article. In a short « *Note on the work of A. Friedmann 'On the curvature of space'* », he argued that *« the results concerning the non-stationary world, contained in [Friedmann's] work, appear to me suspicious. In reality it turns out that the solution given in it does not satisfy the field equations. »*

Of course Friedmann was disappointed. As he could not leave Soviet Union to meet Einstein in Berlin, he wrote an explanatory letter and asked his friend Yuri Krutkov to convince the famous physicist. The mission was apparently successful, since in 1923 Einstein published a still shorter « *Note on the work of A. Friedmann 'On the Curvature of Space'* », where he recongized : « In my previous note I criticised [Friedmann's work On the curvature of Space]. However, my criticism, as I became convinced by Friedmann's letter communicated to me by Mr Krutkov, was based on an error in my calculations. I consider that Mr Friedmann's results are correct and shed new light. They show that the field equations admit, for the structure of spherically symmetric space, in addition to static solutions, dynamical solutions.»

The statement did not mean that Einstein accepted the physical pertinence of dynamical solutions. Indeed, one can find in the original manuscript sent to *Zeitschrift für Physik* that the last sentence did not end as in the published version, but with a concluding disavowal « … to which it is hardly possible to give a physical meaning » !

In 1924, Friedmann studied *On the possibility of a world with constant negative curvature*. He thus assumed a negatively curved space and time varying matter density, and derived the « open universe model », *i.e.* with a dynamics of perpetual expansion.

At the end of this article he had the first insight on a possibly non trivial topology of space : « The knowledge we have about the spatial curvature gives as yet no direct hint about the finiteness or infitineness. To definitely decide about its finiteness one needs additional conditions. As a criterion for the distinctness of points we may take the principle that two points through which more than one geodesic can be drawn are not distinct. It is clear that this principle allows the possibility of « ghosts »: objects and their own images occuring at the same point. This formulation of the sameness and distinctness of points implies that a space of positive curvature is always finite. However it does not allow us to settle the question of the finiteness of a space of negative curvature. »

Unfortunately, the article remained unnoticed, and Friedmann could never gain the satisfaction to see his theoretical models confronted to cosmological observations : he died prematurely in 1925 after an ascent on a balloon (he was also a meteorologist). It was precisely the time when the experimental data began to put in question the validity of static cosmological models. For instance, in 1924 the British theorist Arthur Eddington pointed out that, among the 41 spectral shifts of galaxies as measured by Vesto Slipher, 36 were redshifted ; he thus favored the de Sitter cosmological solution while, in 1925, his PhD student, the young Belgian priest Georges Lemaître, proved a linear relation distance-redshift in de Sitter's solution. The same year 1925, Edwin Hubble proved the extragalactic nature of spiral nebulae. In other words, he confirmed that there existed other galaxies like our own, and the observable Universe was larger than previously expected. More important, the radiation from the faraway galaxies was systematically redshifted, which, interpreted as a Doppler effect, suggested that they were moving away from us at great speed. How was it possible ?

It was Lemaître who solved the puzzle. In his 1927 seminal paper *Un univers homogène de masse constante et de rayon croissant, rendant compte de la vitesse radiale des nébuleuses extragalactiques*, published in French in the *Annales de la Société Scientifique de Bruxelles*, Lemaître calculated the exact solutions of Einstein's equations by assuming a positively curved space (with elliptic topology), time varying matter density and pressure, and a non-zero cosmological constant. He obtained a model with perpetual accelerated expansion, in which he adjusted the value of the cosmological constant such as the radius of the hyperspherical space $R(t)$ constantly increased from the radius of the Einstein's static hypersphere $R_E$ at $t = -\infty$. Therefore there was no past singularity and no « age problem ». The great novelty was that Lemaître provided the first interpretation of cosmological redshifts in terms of space expansion, instead of a real motion of galaxies : space was constantly expanding and consequently increased the apparent separations between galaxies. This idea proved to be one of the most significant discoveries of the century.

Using the available astronomical data, Lemaître provided the explicit relation of proportionality between the apparent recession velocity and the distance : « Utilisant les 42 nébuleuses extra-galactiques figurant dans les listes de Hubble et de Strömberg, et tenant compte de la vitesse propre du Soleil, on trouve une distance moyenne de 0,95 millions de parsecs et une vitesse radiale de 600 km/s, soit 625 km/s à $10^6$ parsecs. Nous adopterons donc R'/R = v/rc = $0,68 \times 10^{-27}$ cm$^{-1}$ (Eq. 24) ». Eq. 24 is exactly what would be called later the Hubble's law !

The fundamental significance of Lemaître's work remained unnoticed. When Lemaître met Einstein for the first time at the 1927 Solvay Conference, the famous physicist told him : « Your calculations are correct, but your physical insight is abominable » !

In 1929, H. P. Robertson mathematically derived the metrics for all spatially homogeneous universes, but did not realize their physical meaning (his compatriot A. Walker did the same job in 1936, so that in the United States, the Friedmann-Lemaître solutions became unfairly called the Robertson-Walker models !)

In 1929, Hubble published the experimental data showing a linear velocity-distance relation *v = Hr* with *H = 600 km/s/Mpc*. This law was strictly identical to Lemaître's Eq.24, with the same proportionality factor … but Hubble did not make the link with expanding universe models. In fact Hubble never read the Lemaître's paper ; he interpreted the galaxy redshifts as a pure Doppler effect (due to a proper velocity of galaxies,) instead of as an effect of space expansion. However, over the course of the 1920's, spiral galaxies were discovered with redshifts greater than 0,1, which implied recession velocities as large as 30,000 km/s! In 1931, in a letter to de Sitter, Hubble expressed his unability to find a theoretical explanation : « We use the term 'apparent velocities' in order to emphasize the empirical features of the correlation. The interpretation, we feel, should be left to you and the very few others who are competent to discuss the matter with authority. » Also he was not aware that the proportionality factor between redshift and distance, wrongly named the « Hubble constant », was not a constant since it varies with time. Thus it is quite erroneous to claim, as it is often the case, that Hubble is the « father » of the big bang theory. In his popular book *The realm of nebulae* (1936), the great astronomer honestly recognizes that « the present author is chiefly an observer» and, on the 202 pages of the book, the theoretical interpretation of observations fills only one page (p. 198) ! Hubble makes reference to Friedmann, Robertson and Milne (who tried to build a newtonian, non relativistic cosmology), but not to Lemaître…

In 1930, Eddington reexamined Einstein's static model and discovered that, like a pen balanced on its point, it is unstable : with the least perturbation, it begins either expanding or contracting. Thus he called for new searches in order to explain the recession velocities in terms of dynamical space models. Lemaître recalled him that he had already solved the problem in his 1927 article. Eddington, who had not read the paper at the right time, made apologies and promoted the Lemaître's model of expanding space. Indeed he translated himself the Lemaître's French original into English for publication in the *Monthly Notices of the Royal Astronomical Society* (1931). Here takes place a curious episode : for an unexplained reason, Eddington replaced the paragraph cited above (where Lemaître gave the relation of proportionality between the recession velocity and the distance) by : « From a discussion of available data, we adopt R'/R = 0,68×10$^{-27}$ cm$^{-1}$ (Eq. 24) ». Thus, due to Eddington's (deliberate ?) blunder, Lemaître will never be recognized on the same foot as Edwin Hubble for being the discoverer of space expansion !

*4. The Primeval Atom*

The same year when his previous work began to be accepted by the scientific community, Lemaître dared to make an even more outrageous assumption: if the universe is expanding now, must it not have been much smaller and denser at some time in the past? In *The Expanding Universe* (*M.N.R.A.S.,* march 1931), he assumed a positively curved space (with elliptic topology), time-varying matter density and pressure, and a cosmological constant such that, starting from a singularity, the Universe first expands, then passes through a phase of « stagnation » during which its radius coasts that of the Einstein's static solution, then starts again in accelerated expansion. This « hesitating model » solved the age problem and provided enough time to form galaxies : « I am led to come around to a solution of the equation by Friedmann where the radius of space starts from zero with an infinite speed, slows and passes by the unstable equilibrium [. . . ] before expanding once again at accelerated speed. It is this period of slowing which seems to me to have played one of the most important roles in the formation of the galaxies and stars. It is obviously essentially connected to the cosmological constant ». Lemaître introduced the revolutionary concept of the « Primeval Atom » : in the distant past the universe must have been so condensed that it was a single entity, which he envisaged as a « quantum of pure energy », referring to the then new discipline of quantum physics. And he poetically described the birth of the Universe : « The atom-world was broken into fragments, each fragment into still smaller pieces […] The evolution of the world can be compared to a display of fireworks that has just ended: some few red wisps, ashes and smoke. Standing on a cooled cinder, we see the slow fading of the suns, and we try to recall the vanishing brilliance of the origin of the worlds . »

In this Lemaître's « annus mirabilis », the short note *The beginning of the world from the point of view of quantum theory,* published in *Nature,* can be considered as the chart of the modern big bang theory. Trying to find a link between nebulae and atoms, he applied the latest knowledge about particles and radioactivity: « A comprehensive history of the universe ought to describe atoms in the same way as stars. […] In atomic processes, the notions of space and time are no more than statistical notions : they fade out when applied to individual phenomena involving but a small number of quanta. If the world has begun with a single quantum, the notions of space and time would altogether fail to have any sense at the beginning and would only begin to get some sensible meaning when the original quantum would have been divided in a sufficient number of quanta. If this suggestion is correct, the beginning of the world happened a little before the beginning of space and time. Such a

beginning of the world is far enough from the present order of nature to be not at all repugnant. »

This idea was very poorly received by other scientists. The fact that Lemaître was a mathematician, allied to his religious convictions (he had been ordained as a priest in 1923), no doubt added to their natural resistance towards the instigation of a new world view. According to Eddington, « the notion of a beginning of the world is repugnant to me », while Einstein considered the primeval atom hypothesis « inspired by the Christian dogma of creation, and totally unjustified from the physical point of view ».

This was an unfair prejudice, because for Lemaître, as he expressed several times, the physical *beginning* of the world was quite different from the metaphysical notion of *creation*. And for the priest-physicist, science and religion corresponded to separate levels of understanding…

Einstein had also a bad opinion of the cosmological constant, that he considered as the « greatest blunder of his life ». It is probably the reason why, in the new relativistic model that he proposed in 1932 with de Sitter – a Euclidean model with uniform density that expanded eternally – the term disappeared. The authors did not even make reference to Friedmann and Lemaître's works, and after that, Einstein forgave research in cosmology…

Unfortunately, due to Einstein's authority, this over-simplified solution became the « standard model » of cosmology for the next 60 years. However Lemaître kept his original views. In *Evolution of the expanding universe*, published in the *Proceedings of the National Academy of Science, USA (1934),* he had a first intuition of a cosmic microwave background as a fossil radiation from the primeval atom : « If all the atoms of the stars were equally distributed through space there would be about one atom per cubic yard, or the total energy would be that of an equilibrium radiation at the temperature of liquid hydrogen. »

He also interpreted for the first time the cosmological constant as vacuum energy : « The theory of relativity suggests that, when we identify gravitational mass and energy, we have to introduce a constant. Everything happens as though the energy in vacuo would be different from zero. In order that motion relative to vacuum may not be detected, we must associate a pressure $p = -\rho c^2$ to the density of energy $\rho c^2$ of vacuum. This is essentially the meaning of the cosmological constant $\lambda$ which corresponds to a negative density of vacuum $\rho_0$ according to $\rho_0 = \frac{\lambda c^2}{4\pi G} \cong 10^{-27} gr./cm.^3$ »

Such a result will be rediscovered only in 1967 by Y. B. Zeldovich on the basis of quantum field theory, and is now considered as one of the major solutions of the so-called « dark energy problem ».

## 5. The hot big bang model

By 1950, when Lemaître published a summary, in English, of his theory, entitled *The Primeval Atom: An Essay on Cosmogony,* it was thoroughly unfashionable. Two years previously the rival theory of a « steady state » universe, supported principally by Thomas Gold in America and by Hermann Bondi and Fred Hoyle in Britain, had met with widespread acclaim. Their argument was that the universe had always been and would always be as it is now, that is was eternal and unchanging. In order to obtain what they wanted, they assumed an infinite Euclidean space, filled with a matter density constant in space and time, and a new « creation field » with negative energy, allowing for particles to appear spontaneously from the void in order to compensate the dilution due to expansion ! Seldom charitable towards his scientific adversaries, Fred Hoyle made fun of Lemaître by calling him « the big bang man ». In fact he used for the first time the expression « big bang » in 1948, during a radio interview.

The term, isolated from its pejorative context, became part of scientific parlance thanks to a Russian-born American physicist George Gamow, a former student of Alexander Friedmann. Hoyle therefore unwittingly played a major part in popularising a theory he did not believe in; he even brought grist to the mill of big bang theory by helping to resolve the question why the universe contained so many chemical elements. Claiming that *all* the chemical elements were formed in stellar furnaces, he was contradicted by Gamow and his collaborators Ralph Alpher and Robert Hermann. The latter took advantage of the fact that the early universe should have been very hot. Assuming a primitive mixture of nuclear particles called *Ylem*, a Hebrew term referring to a primitive substance from which the elements are supposed to have been formed, they were able to explain the genesis of the lightest nuclei (deuterium, helium, and lithium) during the first three minutes of the Universe, at an epoch when the cosmic temperature reached 10 billion degrees. Next they predicted that, at a later epoch, when the Universe had cooled to a few thousand degrees, it suddenly became transparent and allowed light to escape for the first time. Alpher and Hermann calculated that one should today receive an echo of the big bang in the form of « blackbody » radiation at a fossil temperature of about 5 K. Their prediction did not cause any excitement. They refined their calculations several times until 1956, without causing any more interest; no specific attempt at detection was undertaken.

In the middle of the 1960's, at Princeton University, the theorists Robert Dicke and James Peebles studied oscillatory universe models in which a closed universe in expansion-contraction, instead of being infinitely crushed in a big crunch, passes through a minimum radius before bouncing into a new cycle. They calculated that such a hot bounce would cause blackbody radiation detectable today at a temperature of 10 K. It was then that they learned that radiation of this type had just been detected, at the Bell Company laboratories in New Jersey. There, the engineers Arno Penzias and Robert Wilson had been putting the finishing touches on a radiometer dedicated to astronomy, and they had found a background noise which was higher than expected. After subtracting the antenna noise and absorption by the atmosphere, there remained an excess of 3.5 K. This background noise had to be of *cosmic* origin : it was the fossil radiation. The teams of the Bell Company and Princeton University published their articles separately in the same issue of the *Astrophysical journal* (July 1965).

Penzias and Wilson only gave the results of their measurements, while Dicke, Peebles, Peter Roll and David Wilkinson gave their cosmological interpretation. None of them mentioned the predictions of Alpher and Hermann, still less those of Lemaître. The latter died in 1966, a few weeks after his assistant informed him about the discovery of the fossil radiation (Lemaître is supposed to have commented « I am glad now, we have the proof »). Gamow also died in 1968 without being recognized for his predictions. Alpher and Herman were almost forgotten. Penzias and Wilson gained the Nobel prize in physics in 1978. Nevertheless, at the moment of their discovery, they believed instead in the theory of continuous creation, rival to that of the big bang, while their detection of the fossil radiation practically signaled the death sentence of the steady state model !

After half a century of rejection, Lemaître's primeval atom, in the guise of the catchphrase « big bang theory », had at last been accepted by theoretical physicists.

### 6. *The new cosmology*

In 1980, François Englert and Alan Guth independently proposed the theory of inflation, a putative phase of exponential expansion that took place in the very early universe, may be due to the decoupling of fundamental interactions. Although lacking of any observational support and founded on quite uncertain theoretical ground, the inflationary theory became a new « paradigm » of cosmology. It is hardly refutable because, due its large number of degrees of freedom, it generates as many models as necessary to explain this or that cosmological detail.

In 1992, the COBE satellite checked the blackbody nature of the Cosmic Microwave

Background, its homogeneity and isotropy up to $10^{-5}$, and detected the first density fluctuations (G. Smoot and J. Mather, the two principal investigators, won the 2006 Nobel Prize of Physics). Some comments which followed the discovery unhappily forgot the lessons from Lemaître, who had warned against any attempt of concordism between science and religion. I was personally irritated to read the media declarations of cosmologists such as « If you are a believer, it is as if you are looking at the face of God », « They have found the Holy Grail of cosmology », « The greatest discovery of the century, if not all times. »

In 1998 came the first experimental evidence for accelerated expansion. This just signalled that the cosmological constant was back (however renamed « dark energy » to allow for a possible time variation of this repulsive energy field).

Eventually, between 2003 and 2006, the WMAP satellite could survey the CMB with unprecedented precision and allowed us to fix the cosmological parameters within a few per cent. The so-called « concordance » cosmological model (where the CMB observations are combined to other astronomical data) suggests that the age of the universe is 13.7 billion years, that the space curvature is close to zero but probably positive, and that the total energy density is today composed of 28% of gravitating matter (baryonic, non-baryonic, radiation) and 72% of repulsive dark energy. As a consequence, the ultimate fate of our universe would be accelerated perpetual expansion.

Beyond the concordance model, there are intriguing features in the CMB data which do not fit with the usual assumption of an infinite flat space initially driven by inflation. Indeed, power spectrum anomalies suggest a finite, spherical space with dodecahedral topology (Luminet et al., 2003).

*Conclusion*

Big bang models are based on observations and experiments whose results have been extrapolated as far as possible into the past (it is not possible to get back to the very beginning of the universe) and are constructed by a process of hypothesis and calculation -as is the rule in physics. No other kind of model corroborates as many observed phenomena as big bang theory. The latter, which is now almost universally accepted by astrophysicists, can satisfactorily explain the mass of observations made by the great telescopes and the results of experiments carried out in particle accelerators and retrace the principal stages in the creation of the universe – a process which took not six days, but 14 billion years!

It is useful to emphasize that the big bang theory allows for many possible models (depending on cosmological parameters such as the space curvature, the ordinary matter density, the cosmological constant or dark energy field, the space topology, etc.). Some of them are now excluded by experimental data (for instance the strictly flat universe in decelerated expansion filled in only with ordinary matter, namely the Einstein-de Sitter model, much in favour in the 1930-1980's), but the general picture (*i.e.* a presently universe starting from an initial hot dense configuration) is much reinforced.

A confusion sometimes present in the mind of some cosmologists is that the big bang theory is synonymous of inflation theory. However, the latter – or at least the usual inflationary models – is seriously challenged by WMAP data when one looks at the power spectrum « anomalies » . Of course the « conservative cosmologists » prefer to consider that the anomalies are artifacts, coming from bad data analysis. Among those researchers who believe that the anomalies are reliable, some inflationists invoke a special feature in the inflaton field, using the well-known theorem « inflation can do everything ». However, adding special features and fitting free parameters in the speculative inflation theory to « save the apparences » looks much like adding epicycles in Ptolemy's theory.  There is no physical model behind this!

Eventually, some cosmologists think that anomalies are reliable, some of them (the low quadrupole and octopole) having a geometrical explanation in terms of a finite space with a non trivial topology, while others anomalies (violations of statistical isotropy in the same multipoles) are due to local effects. My belief is clearly this one (Luminet, 2005).

It would not be a fundamental upheaval of relativistic cosmology to modify, or even abandon the inflation scenario. It would be more interesting (in my opinion) to get a definite clue of the finiteness and the non trivial topology of space – but, as said above, all this is already potentially included in the large family of big bang solutions. A major upheaval would rather be related to the confirmation of more radical new views which, in the framework of quantum gravity theories such as superstrings, M-theory or quantum loop gravity (see e.g. Smolin, 2002), allow for entirely new phenomena, e.g. additional space dimensions, pre-big bang models, multiverse, etc. This would really change the present-day cosmological « paradigm ». Neither WMAP or Planck Surveyor satellites will do that. May be some high energy experiments in particle accelerators will provide hints for a drastically new vision of the Universe we inhabit.